\documentclass[doublecol,linenumbers]{epl2} 

\usepackage{color}

\title{Long-ranged velocity correlations in dense systems of self-propelled particles}
\shorttitle{Long-ranged velocity correlations} 

\author{Grzegorz Szamel\thanks{E-mail: \email{grzegorz.szamel@colostate.edu}} \and Elijah Flenner}
\shortauthor{G. Szamel \etal}

\institute{Department of Chemistry,
Colorado State University, Fort Collins, CO 80523, USA
}
\pacs{05.40.-a}{Fluctuation phenomena, random processes, noise and Brownian motion}
\pacs{82.70.Dd}{Colloids}
\pacs{47.57.-s}{Complex fluids and colloidal systems}

\abstract{
Model systems of self-propelled particles reproduce many phenomena observed
in laboratory active matter systems that defy our thermal equilibrium-based intuition. In particular, 
in stationary states of self-propelled systems, it is 
recognized that velocities of different particles exhibit 
non-trivial equal-time correlations. Such correlations are absent in
equivalent equilibrium systems. Recently, researchers found that the range of the velocity correlations increases with increasing 
persistence time of the self-propulsion and can extend over many particle diameters. 
Here we review the initial studies of long-ranged velocity correlations in solid-like 
systems of self-propelled particles. Then, we demonstrate that the long-ranged velocity correlations are also
present in dense fluid-like systems. We show that the range of velocity correlations in dense systems of self-propelled particles
is determined by the combination of the self-propulsion and the virial bulk modulus that originates from 
repulsive interparticle interactions.}

\begin{document}

\maketitle

\section{Introduction }
A quickly growing field is the study of active matter systems 
\cite{Ramaswamy2010,Marchetti2013,Vicsek2012,Bechinger2016,Elgeti2015,Needleman2017,Gompper2020}. 
Individual components of these systems perform persistent motion due to the injection (consumption) of energy from their environment. 
Examples include cell assemblies \cite{Petitjean2010,Angelini2011,Basan2013,Garcia2015,Blanch2018,Henkes2020}, 
bird flocks \cite{Vicsek2012}, bacterial suspensions 
\cite{Dombrowski2004,Peruani2012,Wensink2012,Dunkel2013,Wioland2016,Urzay2017,James2018}, 
and self-propelled colloids \cite{Howse2007,Tierno2008,Gosh2009,Palacci2010,Jiang2010,Michelin2013,Dai2016,Moran2017}. 
Active matter systems exhibit many properties absent in equilibrium thermal systems, \textit{e.g.} they may undergo a phase
separation of liquid-gas type in the absence of any attractive interactions \cite{Cates2015}. 

One interesting property, first demonstrated experimentally by Garcia \textit{et al.} \cite{Garcia2015}, 
is the presence of equal-time velocity
correlations. These correlations are absent in classical equilibrium systems. 
It has been recognized for some time \cite{Szamel2015,Marconi2016,Flenner2016} that such non-trivial equal-time velocity correlations 
are present in simple microscopic models of active matter systems, \textit{i.e.} in systems of self-propelled particles. 
These correlations are an \textit{emergent property} of these systems, \textit{i.e.} they appear spontaneously, 
without any explicit velocity-aligning interactions. 
Recently, two groups \cite{Henkes2020,Caprini2020a} independently found that velocity correlations in dense systems of 
self-propelled particles can be long-ranged\footnote{Long range of velocity correlations was noted in passing in early work 
\cite{Flenner2016}, but it was not studied systematically.}. The analysis and rationalization of these correlations relied upon 
the solid-like nature of the systems studied. Here we review these studies and present computer simulation results that 
demonstrate the presence of long-ranged velocity correlations also in dense fluid-like systems. We develop a simple theory that explains
the appearance of these correlations through the combined effect of the self-propulsion and the virial bulk modulus of the active fluid.
We finish with a brief discussion, emphasizing the features of velocity correlations that the simple theory cannot describe. 

\section{Long-ranged velocity correlations in dense, ordered systems of self-propelled particles}
Caprini \textit{et al.} \cite{Caprini2020a,Caprini2020b} investigated $2d$ systems of 
repulsive, monodisperse, overdamped active Brownian particles (ABPs) \cite{tenHagen,FilyMarchetti}. 
They showed that the previously studied motility-induced phase separation \cite{Cates2015}
is accompanied by a spontaneous alignment of the velocities of the particles in the dense phase. 
They found that the dense phase is either hexatic or solid, and that the transition between these two phases influences the alignment. 
The average size of the domains with aligned velocities was found to grow with increasing
persistence time of the self-propulsion. Although the spontaneous alignment of the velocities was mainly discussed in the context of 
phase separation, Caprini \textit{et al.} showed that it also occurred in single-phase systems if the density was 
high enough. While sufficiently high density was important for the appearance of the spontaneous velocity ordering, the size of the
ordered domains was growing primarily due to increasing persistence time. 

To quantify the observed velocity ordering Caprini \textit{et al.} introduced and evaluated two correlation functions. 
Here we focus on velocity correlation function $C(r)$ \cite{Caprini2020b},
\begin{equation}\label{Crdef}
C(r) = \frac{\left<\mathbf{v}(\mathbf{r})\cdot\mathbf{v}(0)\right>}{\left<v^2\right>},
\end{equation}
where $\mathbf{v}(\mathbf{r})$ represents the velocity of the particle located at $\mathbf{r}$ (continuous limit is implied).
Caprini \textit{et al.} found that in dense systems $C(r)$ 
exhibits exponential dependence on $r$. 
We note that exponentially decaying velocity correlations were found in the experimental study of Garcia \textit{et al.} \cite{Garcia2015}.
Using $C(r)$, Caprini \textit{et al.} defined and evaluated the velocity correlation length. They found that it 
increases as the square root of the persistence time of the self-propulsion, 
but it is also influenced by the transition between the solid and hexatic phases. 

To explain their findings Caprini \textit{et al.} developed a theory for velocity correlation function $C(r)$ based on the
assumption that the dense phase is a $2d$ hexagonally ordered crystal, with particles oscillating around their average positions. 
While this assumption is appropriate for dense ordered systems investigated in Refs. \cite{Caprini2020a,Caprini2020b},
it is not applicable for fluid-like disordered systems. Caprini \textit{et al.} 
showed that their assumption results in the following formula for the large-$r$ behavior of the correlation function,
\begin{equation}\label{Crth}
C(r) \propto \frac{\bar{x}^2}{\ell^2}\left(\frac{\ell}{8\pi r}\right)^{1/2} e^{-r/l},
\end{equation}
where $\bar{x}$ is the lattice constant and correlation length $\ell$ is given by
\begin{equation}\label{lth}
\ell = \bar{x} \sqrt{\frac{\tau}{\gamma}} \left[\frac{3}{4}\left(U''(\bar{x}) + \frac{U'(\bar{x})}{\bar{x}}\right)\right]^{1/2}.
\end{equation}
In Eq. (\ref{lth}) $\tau$ is the persistence time of the self-propulsion, $\gamma$ is the friction coefficient of an isolated particle
and $U(r)$ is the interparticle interaction potential. 

\begin{figure}
\begin{center}
\includegraphics[width=0.48\columnwidth]{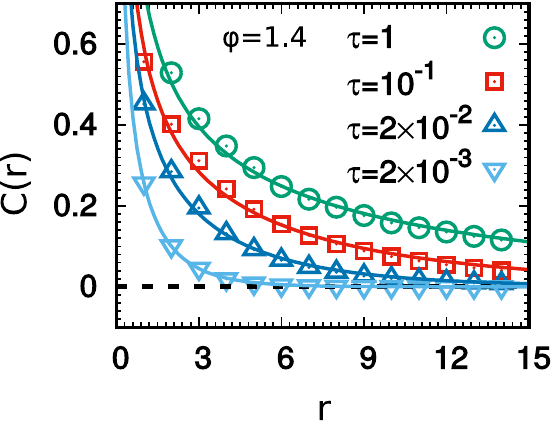} \hskip .2em \includegraphics[width=0.47\columnwidth]{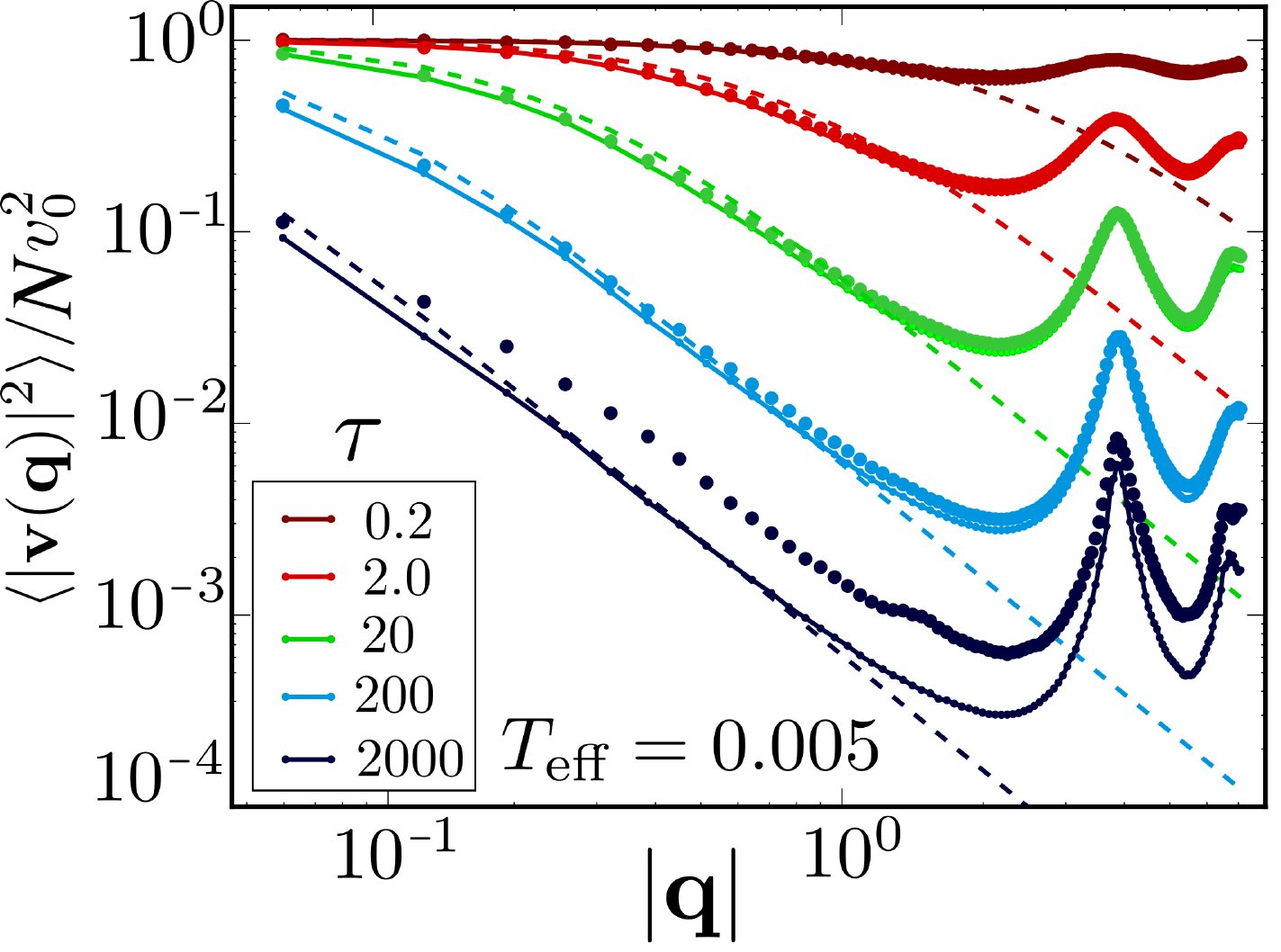}
\end{center}
\caption{\label{CapriniHenkesfig} 
Left panel: velocity correlations for ABP systems studied in Refs. \cite{Caprini2020a,Caprini2020b} 
for different persistence times $\tau$ at $v_0=50$. Symbols: simulation results. Lines: theoretical predictions.
Reprinted with Author's permission from Ref. \cite{Caprini2020b}.
Right panel: velocity correlations for ABP systems studied in Ref. \cite{Henkes2020} for 
different persistence times $\tau$ at $T_a=0.005$. Dots: simulation results. Lines: theoretical predictions; 
solid line: normal mode-based approach; dashed lines: continuum elasticity approach. Reprinted with Author's permission 
from Ref. \cite{Henkes2020}.}
\end{figure}

Caprini \textit{et al.} found that the approach outlined above describes the behavior of velocity correlation function very well, 
see the left panel of Fig. \ref{CapriniHenkesfig} for an example. 

In 
a recent work Caprini and Marconi \cite{Caprini2020c} investigated underdamped 
analogues of active Brownian particles systems. They showed that long-ranged equal-time velocity correlations 
persist in the presence of inertia and 
thermal fluctuations. We note that Caporuso \textit{et al.} 
\cite{Caporusso2020} did not find long-ranged equal-time velocity correlations in overdamped systems of active Brownian particles with 
thermal fluctuations. We suggest that further work is needed to clarify these somewhat conflicting results.

\section{Long-ranged velocity correlations in dense amorphous systems of self-propelled particles}
Henkes \textit{et al.}\ \cite{Henkes2020} completed a combined experimental, simulational and theoretical study
of velocity correlations in active matter systems. On the experimental side they studied the dynamics of epithelial cell
monolayers and found displacement and velocity correlations over several cell sizes. They found that the displacement correlations
resembled those observed in supercooled liquids. Conversely, there are no equal-time velocity correlations in 
liquids.

Henkes \textit{et al.} simulated $2d$ systems of 
repulsive polydisperse overdamped active Brownian particles. They also simulated
systems of polydisperse self-propelled Voronoi cells \cite{Bi2016,Barton2017}, which can be thought of as active objects with complicated 
many-particle (non-pairwise-additive) interactions. The polydispersity was introduced to account for cell size heterogeneity. As a result,
systems simulated by Henkes \textit{et al.} remained amorphous in the range of the parameters used in their study.

Henkes \textit{et al.} found that non-trivial equal-time velocity correlations exist in both simulated systems. In agreement with the 
results obtained by Caprini \textit{et al.}, the range of these correlations was found to increase as the square-root of the 
persistence time of the self-propulsion.  

To explain their results Henkes \textit{et al.} assumed that on short enough time scales, their systems can be approximated
by amorphous elastic solids. They used two related approaches. First, they generated local potential energy minima (inherent structures) 
corresponding to configurations obtained in simulations and approximated the real short-time dynamics by harmonic motion around 
these minima. They found the associated normal modes and expressed the equal-time velocity correlations in terms
of normal mode frequencies, normal mode amplitudes and the persistence time. Second, Henkes \textit{et al.} postulated a continuum elastic 
description of their active  systems. In this case, the equal-time velocity correlations were expressed in terms of the elastic
bulk and shear moduli, and the persistence time,
\begin{equation}\label{vqel}
\left<|\mathbf{v}(\mathbf{q})|^2\right> = \frac{Nv_0^2}{2}
\left[\frac{1}{1+\left(\xi_L q\right)^2}+\frac{1}{1+\left(\xi_T q\right)^2}\right]
\end{equation}
where $\mathbf{v}(\mathbf{q}) = \sum_j \dot{\mathbf{r}}_je^{-i\mathbf{q}\cdot\mathbf{r}_j}$, 
$N$ is the number of particles, $v_0$ is the self-propulsion velocity. The
correlation lengths $\xi_L$ and $\xi_T$ can be expressed in terms of the bulk $B$ and shear $\mu$ moduli as
$\xi_L^2 = \left(B+\mu\right)\tau/\gamma$ and $\xi_T^2 = \mu\tau/\gamma$.

The normal mode-based approach gives quite accurate predictions for both small
and large wavevectors, 
see the solid line in the right panel of Fig. \ref{CapriniHenkesfig}.
By construction, the continuum elastic approach is only applicable in the small wavevector (large distance) limit, and in this 
limit it reproduces the results of the normal mode approach, see the dashed lines in the right panel of Fig. \ref{CapriniHenkesfig}.

While Henkes \textit{et al.}'s approach is generally applicable to active systems exhibiting slow glassy-like dynamics,
from a physical point of view it seems inapplicable to dense active systems with constituents moving perhaps slowly but in a 
standard, fluid-like fashion. 

\section{Long-ranged velocity correlations in dense fluid-like systems of self-propelled particles} 
We originally stumbled upon non-trivial equal-time velocity correlations when developing a theory for the dynamics of 
of active Ornstein-Uhlenbeck particles \cite{Szamel2014,Maggi2015,Fodor2016}. We found that these correlations
determine their short-time dynamics \cite{Szamel2015,Flenner2016,Berthier2017,Berthier2019}
and they also appear in an approximate mode-coupling-like theory for active particle systems \cite{Szamel2016}.

Here we present computer simulation results showing that velocity correlations in dense active fluid-like systems are long-ranged. 
To rationalize this finding we develop
a simple theory similar to the one presented by Henkes \textit{et al.} Our theory does not assume elastic response, is applicable
to fluid-like systems and describes the major part of the long-ranged velocity correlations.

\subsection{Simulations}
We simulated 
two-dimensional 
polydisperse systems of active Brownian particles \cite{tenHagen,FilyMarchetti}. 
The equations of motion for the position $\mathbf{r}_i$ and the angle $\phi_i$ specifying the orientation of the 
self-propulsion of particle $i$ are given by
\begin{eqnarray}\label{rdot}
\gamma \dot{\mathbf{r}}_i &=& - \nabla_i \sum_{j} V(r_{ij}) + \gamma v_0 \mathbf{n}_i 
\\ \label{phidot}
\dot{\phi}_i &=& \eta_i,   
\end{eqnarray}
where $v_0$ is the self-propulsion velocity, $\mathbf{n}_i = (\cos(\phi_i),\sin(\phi_i))$ is the orientation vector 
specifying the direction of the self propulsion, and the random variable $\eta(t)$ satisfies 
$\left< \eta_i(t) \eta_j(t^\prime) \right> = 2 D_r \delta_{ij} \delta(t-t^\prime)$. In $2d$, persistence time of the self-propulsion, 
$\tau$, is the inverse of the rotational diffusion coefficient, $\tau=1/D_r$. For an isolated active particle,
Eqs. (\ref{rdot}-\ref{phidot}) 
result in a mean-square displacement that for long times grows as 
$\left< \delta r^2(t) \right> \simeq \frac{2 v_o^2}{D_r} t$. Comparing this result to the mean-square displacement
of a Brownian particle in $2d$, $\left< \delta r^2(t) \right> = 4 T/\gamma$, we can define an active temperature 
$T_a = v_0^2\gamma/(2 D_r)$.  

The interaction potential is given by  
\begin{equation}
\label{potential}
V(r_{ij}) = \epsilon \left( \frac{\sigma_{ij}}{r_{ij}} \right)^{12} + c_0 
+ c_2 \left(\frac{r_{ij}}{\sigma_{ij}} \right)^2 + c_4 \left(\frac{r_{ij}}{\sigma_{ij}} \right)^4,
\end{equation}
when the distance between particles $i$ and $j$, $r_{ij} < 1.25 \sigma_{ij}$ and zero otherwise. In Equation \ref{potential}
coefficients $c_\alpha$ are chosen so that the potential and the first two derivatives are 
continuous.  The diameters $\sigma_i$ are chosen from the distribution $P(\sigma) = A/\sigma^3$ for $0.73 < \sigma < 1.63$. 
The cross diameter $\sigma_{ij} = 0.5(\sigma_i+\sigma_j)(1-0.2 |\sigma_i-\sigma_j|)$. The number density is $\rho = 1.23$. 
The interaction potential and the density are chosen to prevent crystallization and significant structural changes for a large range of 
simulation parameters.  

Most of the simulations were done using $N=10 000$ particles. 
Due to the long-range of the velocity correlations, the simulations for the three lowest $D_r$ used $N=250 000$ particles.

\begin{figure}
\begin{center}
\includegraphics[width=0.49\columnwidth]{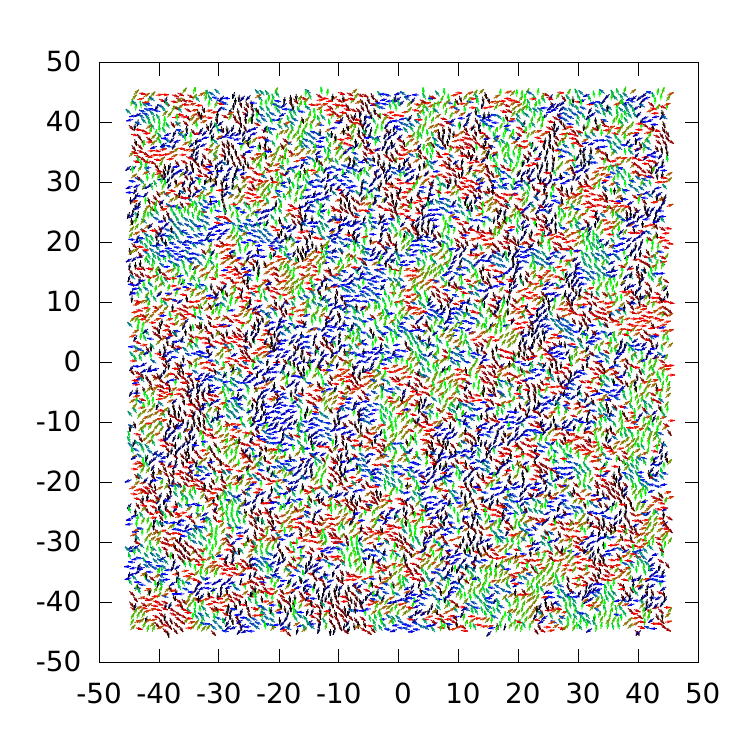} \hskip .2em
\includegraphics[width=0.49\columnwidth]{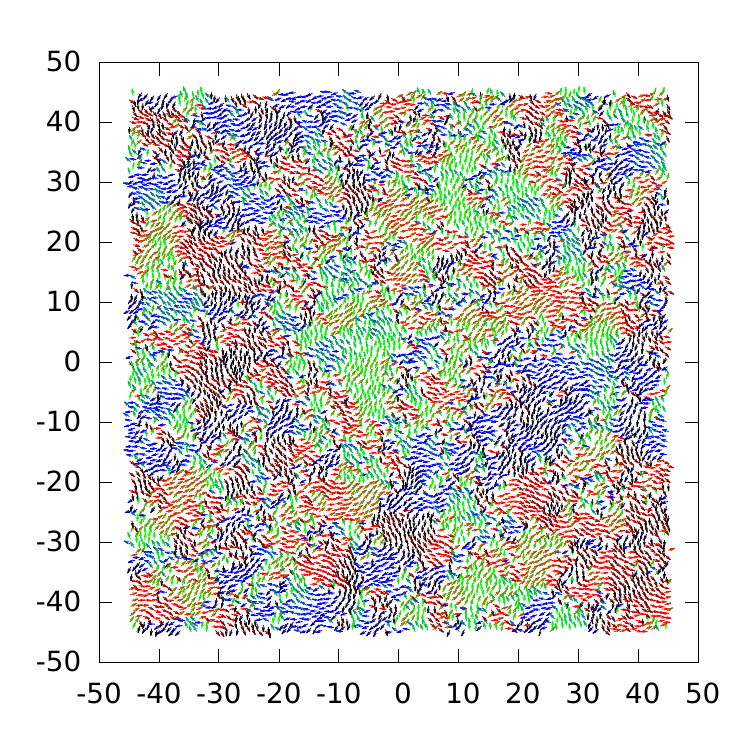}
\end{center}
\caption{\label{visual} 
Snapshots of configurations for $\tau=0.14$ (left panel) and $\tau=10.0$ (right panel). Arrows show
orientations of the velocities; specific velocity directions are also color-coded.}
\end{figure}

\subsection{Velocity Correlations}
Qualitatively, the increase of the range of velocity correlations is evident from snapshots shown in Fig. \ref{visual}. 
To quantify these correlations we introduce two correlation functions. The first function,
\begin{equation}\label{opardef}
\omega_{\parallel}(q) = \frac{1}{N} 
\left< \left| \hat{\mathbf{q}} \cdot \mathbf{v}(\mathbf{q})
\right|^2 \right>,
\end{equation}
which we refer to as the longitudinal velocity correlation function, appeared naturally in the analysis of the 
short-time behavior of the intermediate scattering function of self-propelled particles \cite{Szamel2015}.
Here we examine $\omega_{\parallel}(q)$ and the complementary part of velocity correlations, the transverse velocity correlation function,
\begin{equation}\label{operpdef}
\omega_{\perp}(q) = 
\frac{1}{N} \left< 
\left|\mathbf{v}(\mathbf{q})-\hat{\mathbf{q}}(\hat{\mathbf{q}}\cdot\mathbf{v}(\mathbf{q}))
\right|^2 \right>.
\end{equation}
The wavevectors $\mathbf{q}$ in Eqs. (\ref{opardef}-\ref{operpdef}) have to satisfy periodic boundary conditions.  
Both $\omega_{\parallel}(q)$ and $\omega_{\perp}(q)$ are equal to $v_0^2/2$ for $q = 0$, which is 
useful in fits for the correlation length described below. 

\begin{figure}
\begin{center}
\includegraphics[width=0.8\columnwidth]{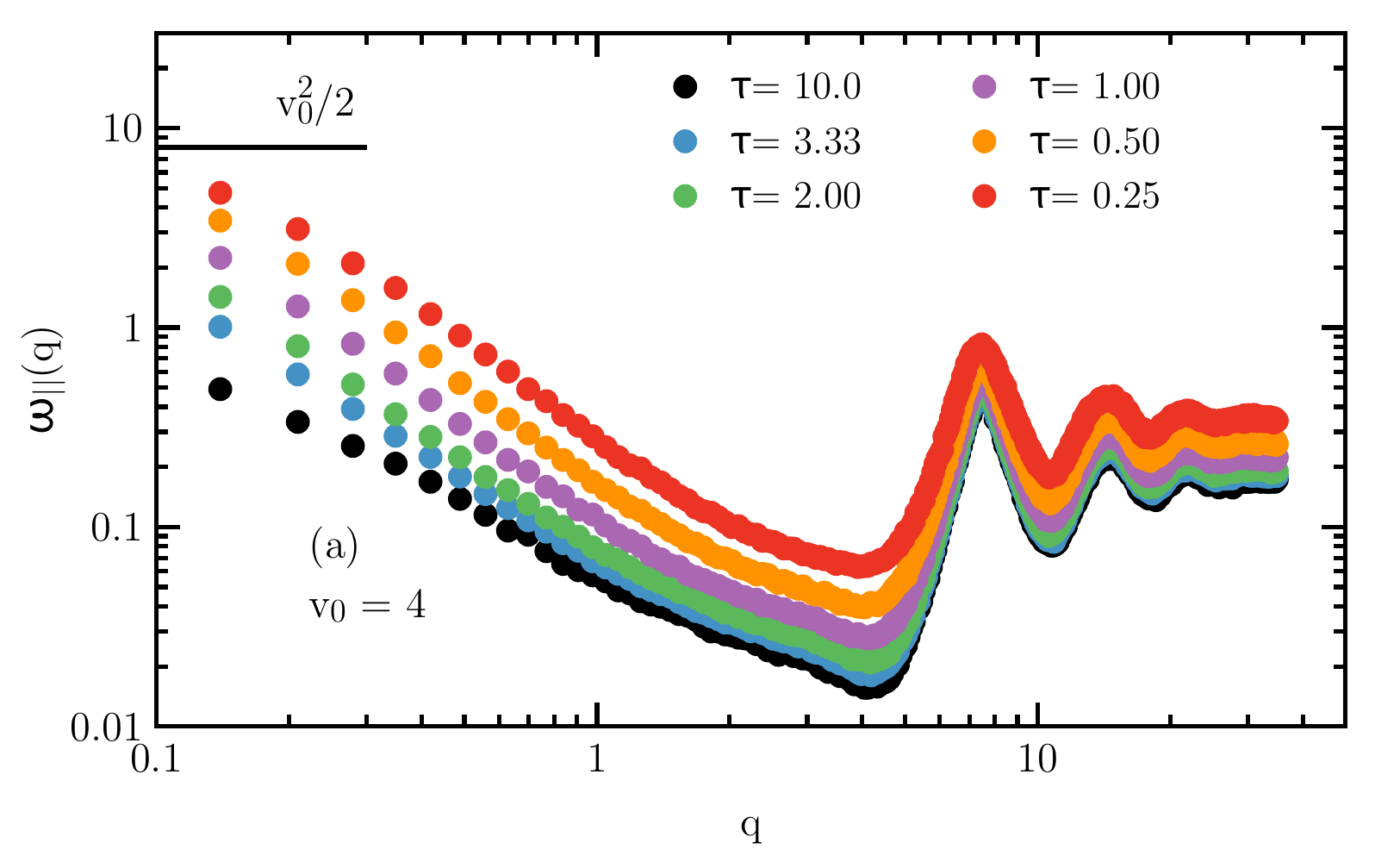}
\includegraphics[width=0.8\columnwidth]{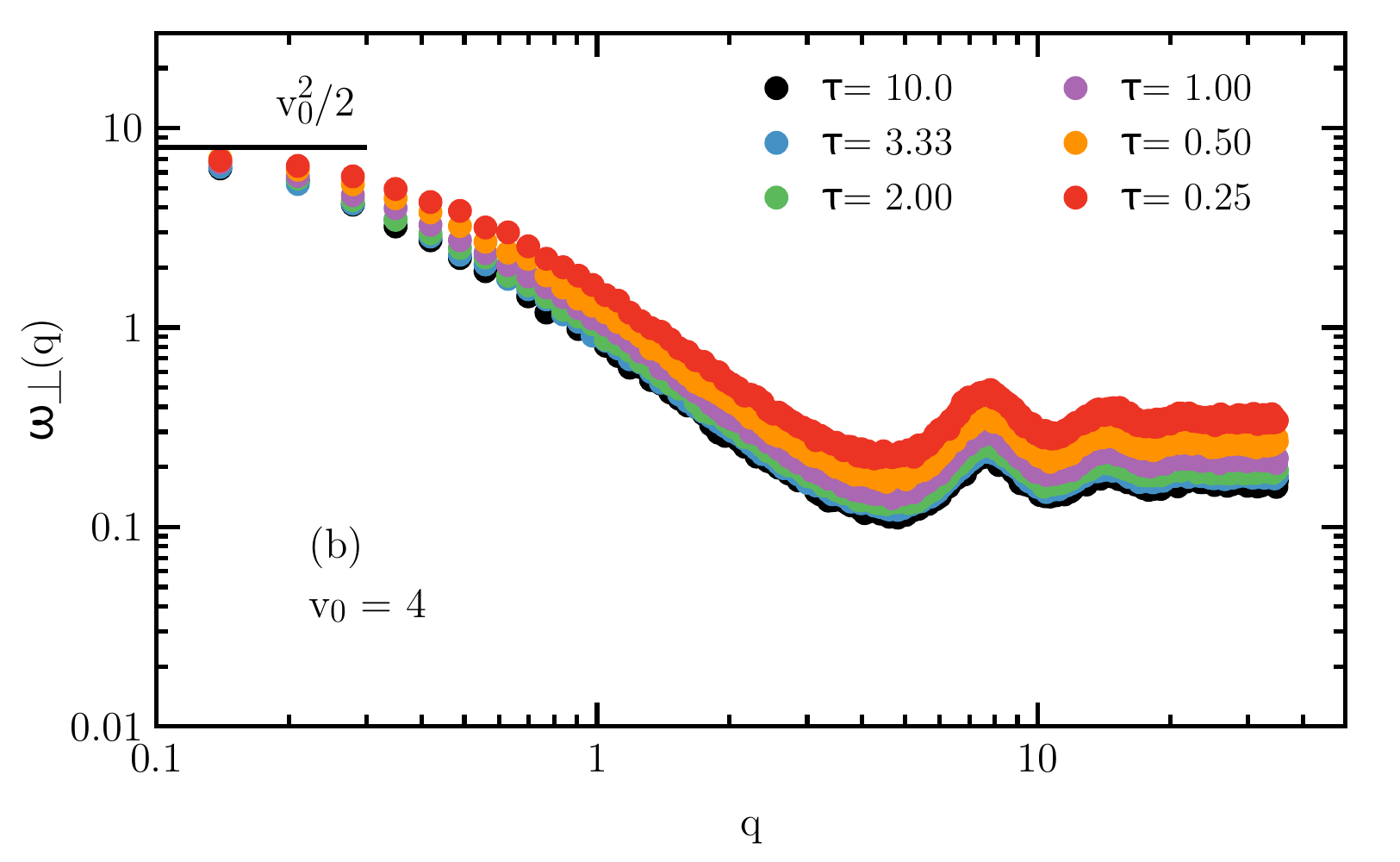}
\end{center}
\caption{\label{omega}The equal time velocity correlation functions (a) $\omega_{\parallel}(q)$ and (b) $\omega_{\perp}(q)$ 
calculated at a fixed $v_0 = 4$ for a range of $\tau$. The longitudinal correlation function $\omega_{\parallel}(q)$ indicates 
a a correlation length rapidly growing with increasing $\tau$. }
\end{figure}

The large number of parameters, $v_0$, $D_r$, $\rho$, necessitates choosing some cuts through the parameter space. 
We note that 
Caprini \textit{et al.} fixed $v_0$ and examined velocity correlations as a function of the persistence time $\tau=1/D_r$ 
for various densities whereas Henkes \textit{et al.} fixed $\rho$ and examined velocity correlations as a function of $\tau$ 
for a fixed value of active temperature $T_a=v_0^2\gamma/D_r$ and as a function of $T_a$ for a fixed value of $\tau$. 
Here we fix $\rho=1.23$ and examine velocity correlations as a function of $\tau=1/D_r$, first for a fixed value of $v_0=4$ and
then for a fixed value of $T_a=8$. One interesting feature of the latter procedure is that the system approaches a Brownian 
system with $T=T_a$ for $\tau\to 0$, and it has been argued that with increasing $\tau$ at fixed $T_a$ the active system
moves systematically farther away from equilibrium \cite{Flenner2020}. 

\begin{figure}
\begin{center}
\includegraphics[width=0.8\columnwidth]{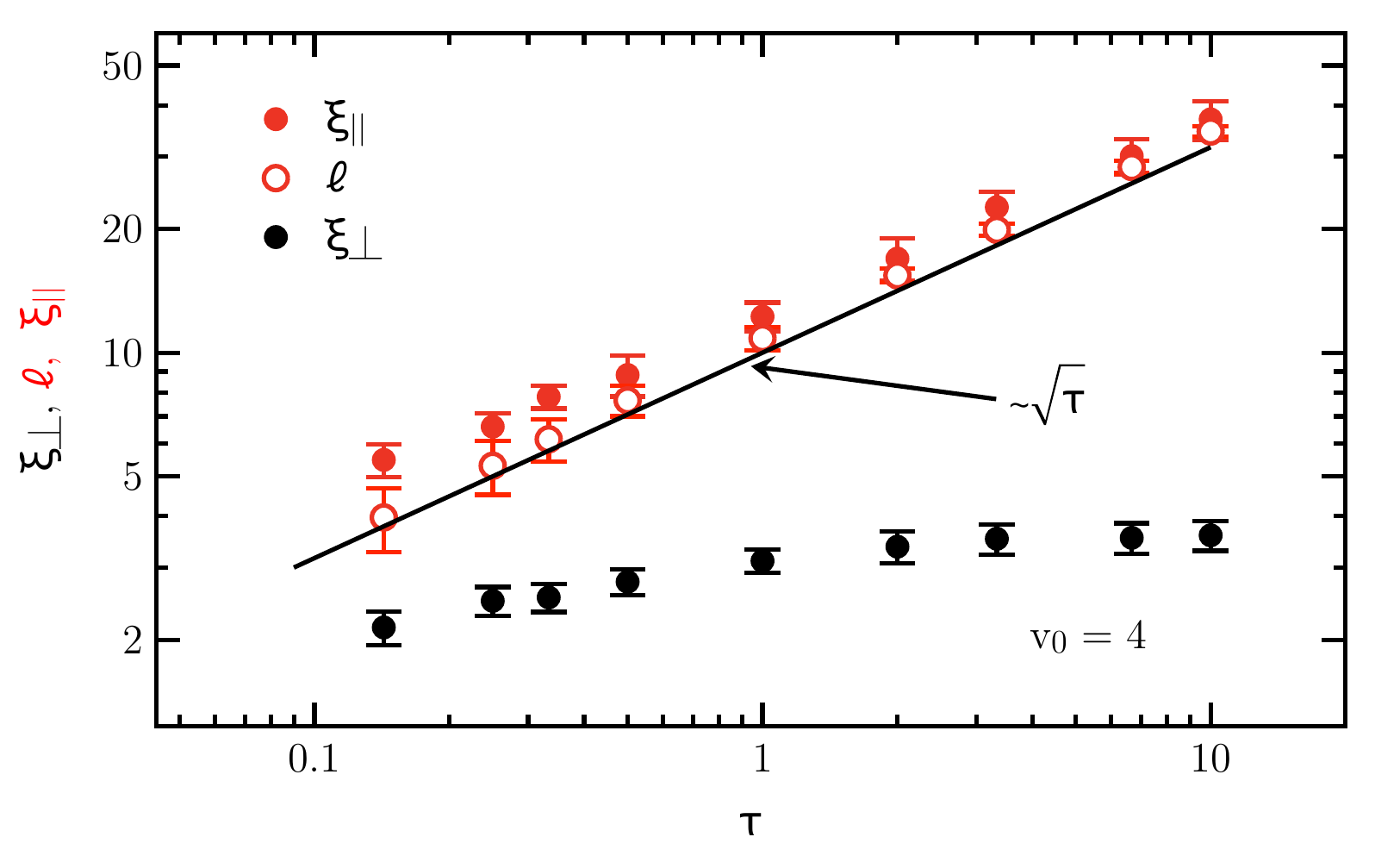}
\end{center}
\caption{\label{length} The longitudinal correlation length $\xi_{\parallel}$ obtained from fitting $\omega_{\parallel}(q)$ 
(filled red circles) and transverse correlation length $\xi_{\perp}$ (filled black circles) 
obtained from fitting $\omega_{\perp}(q)$ (closed black circles) 
for a fixed self-propulsion velocity $v_0=4$. The longitudinal 
correlation length grows approximately as $\sqrt{\tau}$, while the transverse correlation length is almost unchanged
for a fixed $v_0$. The open red circles are results of the 
approximate theory. }
\end{figure}

In Fig.~\ref{omega}(a) we show $\omega_{\parallel}(q)$ for $v_0 = 4$ and a range of $\tau$ from 0.14 to 10.
Since $\omega_{\parallel}(q=0)$ is constant for a fixed $v_0$, it is apparent that there is a faster decay of $\omega_{\parallel}(q)$
at the small wavelengths with increasing $\tau$, which corresponds to a longer range of longitudinal velocity correlations.

In contrast, $\omega_{\perp}(q)$, which is shown in Fig.~\ref{omega}(b) changes little with $\tau$, which implies that 
transverse velocity correlations are significantly less dependent on $\tau$ at fixed $v_0$.

To determine the velocity correlation length, we fitted $\omega_{\alpha}(q)$ for $q < 0.2$ 
to an Ornstein-Zernike-like form $(v_0^2/2)/[1 + (\xi_{\alpha}q)^2] + \omega_{\alpha}(\infty)$. 
In Fig.~\ref{length} we show correlation lengths $\xi_{\parallel}$ and $\xi_{\perp}$ obtained from fitting
$\omega_{\parallel}(q)$ and $\omega_{\perp}(q)$, respectively. The longitudinal correlation length increases from 
approximately 4 at $\tau = 0.14$ to 34 at $\tau = 10.0$ whereas the transverse length 
increases only slightly over the full range of $\tau$.

\begin{figure}
\begin{center}
\includegraphics[width=0.8\columnwidth]{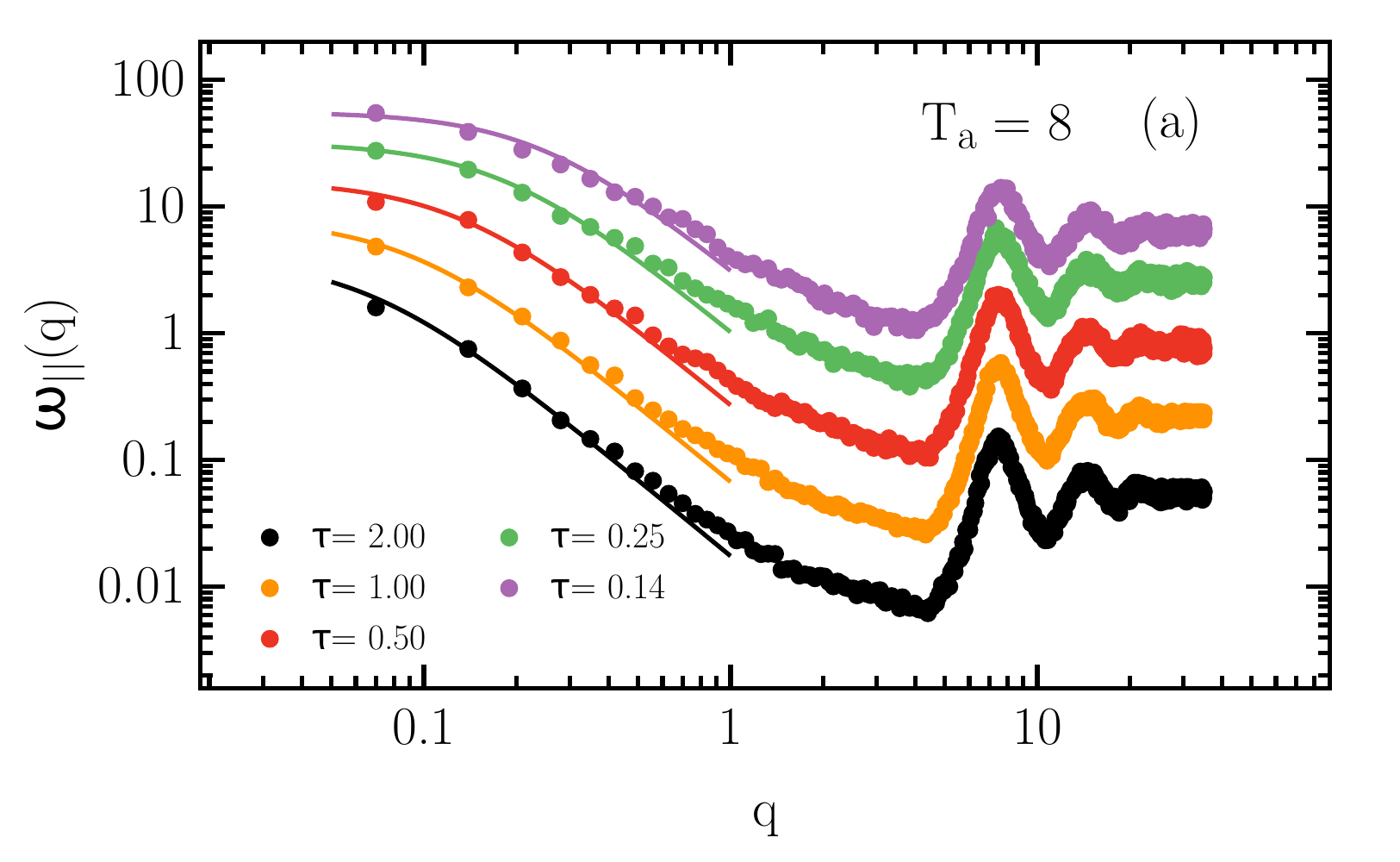}
\includegraphics[width=0.8\columnwidth]{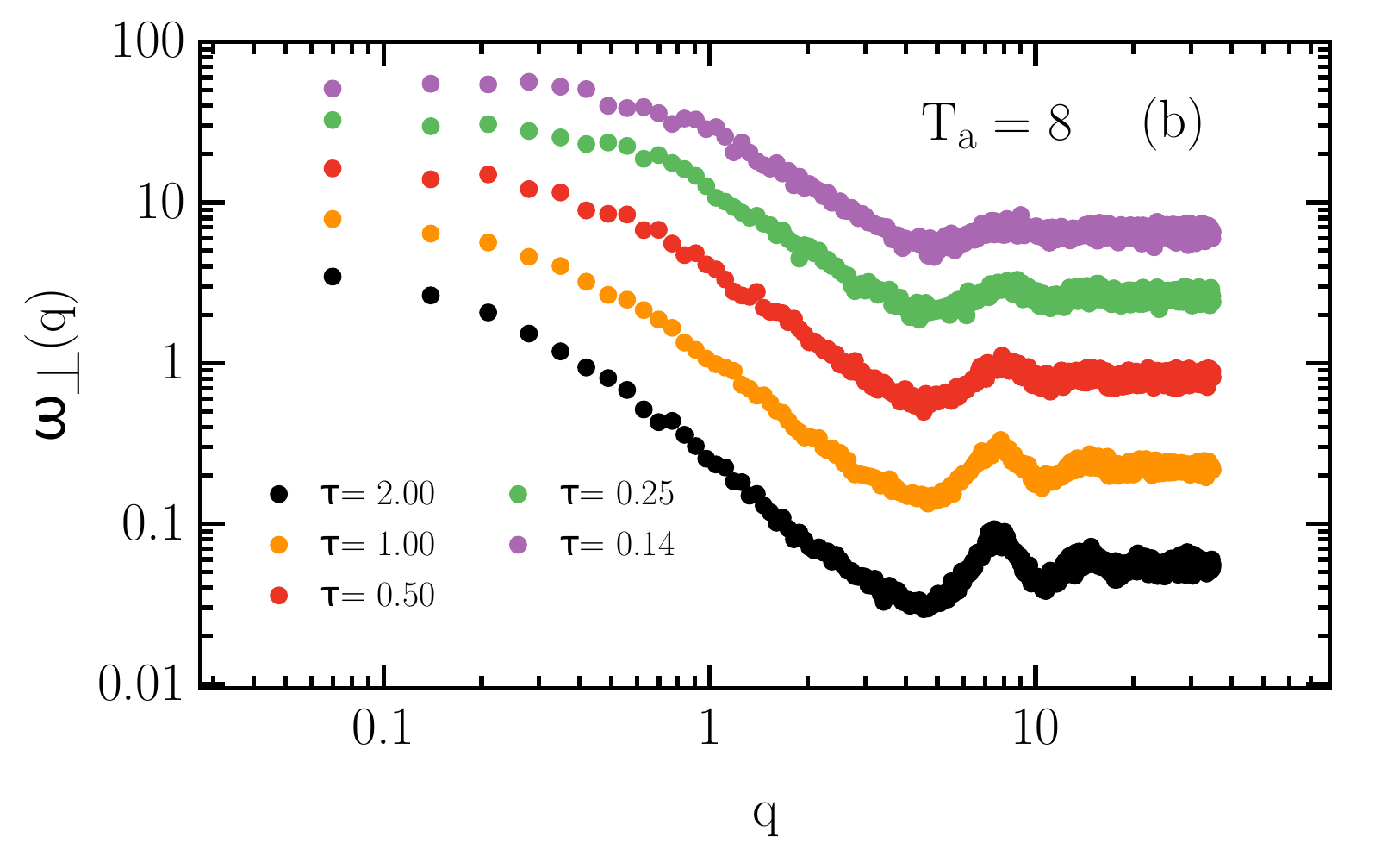}
\end{center}
\caption{\label{omegaT8} The equal time velocity correlation functions (a) $\omega_{\parallel}(q)$ and (b) $\omega_{\perp}(q)$ 
calculated at a fixed active temperature $T_a = 8$ for a range of $\tau$. The longitudinal correlation function $\omega_{\parallel}(q)$ 
indicates a correlation length growing with increasing $\tau$. The transverse correlation function $\omega_{\perp}(q)$ also suggests a 
growing length scale, albeit a smaller one. Solid lines in (a) are predictions of the approximate theory, Eq. (\ref{corrstheory}).}
\end{figure}

In Fig.~\ref{omegaT8} we show $\omega_{\parallel}(q)$ 
and $\omega_{\perp}(q)$ 
for a fixed $T_a = 8$.
Since increasing persistence time for a fixed value of the 
active temperature results in a rapidly slowing diffusive motion of the particles, we could
only simulate a restricted range of $\tau$. Once again we observe increasing range of velocity correlations, but
for fixed $T_a$ the range of both longitudinal and transverse correlations is increasing.  

\begin{figure}
\begin{center}
\includegraphics[width=0.8\columnwidth]{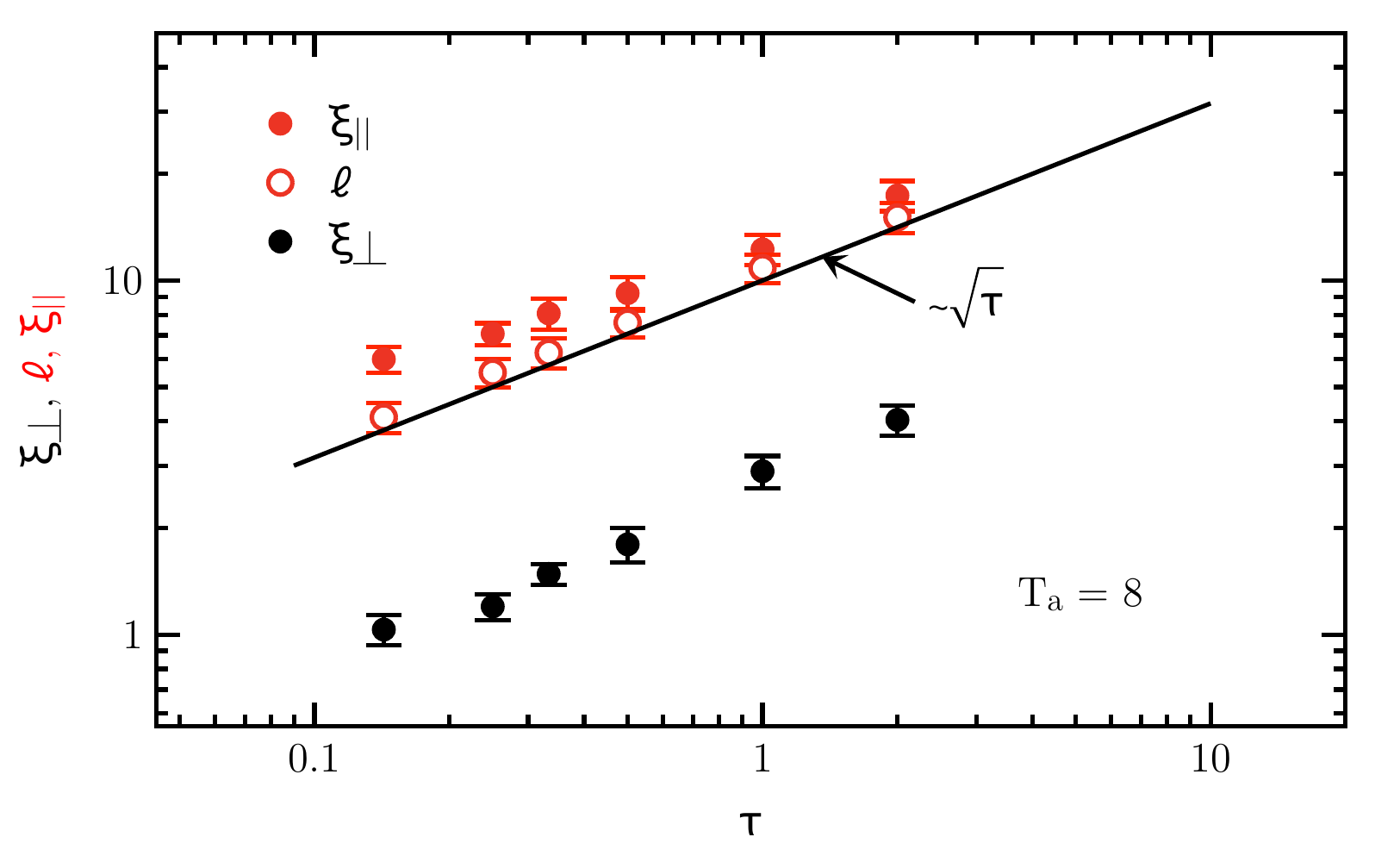}
\end{center}
\caption{\label{lengthT8}The longitudinal velocity correlation length $\xi_{\parallel}$ (filled red circles) and the 
transverse correlation length $\xi_{\perp}$ (filled black circles) for a fixed active temperature $T_a = 8$. Solid lines in (a) represent  
theoretical predictions. The longitudinal correlation
length is much larger than the transverse correlation length, but they both grow approximately as $\sqrt{\tau}$ 
for fixed $T_a$. The open red circles are predictions of the approximate theory.}
\end{figure}

We obtained velocity correlation lengths from fits to an Ornstein-Zernicke-like function and we present these lengths in
Fig.~\ref{lengthT8}. For large $\tau$ both correlation lengths grow with increasing persistence time. We recall that we 
did not observe growing $\xi_{\perp}$ at fixed $v_0$, which implies that this length must also depend on $v_0$.

\subsection{Theory}
\label{theory}
Our starting point for an approximate theory for velocity correlations is equation of motion (\ref{rdot}) from which we 
derive the following relation between velocity, polarization and force fields in the Fourier space,
\begin{equation}\label{motion}
\gamma \mathbf{v}(\mathbf{q};t) = \sum_j \sum_{k \ne j} \mathbf{F}_{jk}e^{-i \mathbf{q} \cdot \mathbf{r}_j}
+\gamma v_0 \mathbf{n}(\mathbf{q};t)
\end{equation} 
where $\mathbf{v}(\mathbf{q};t) = \sum_j \dot{\mathbf{r}}_j e^{-i \mathbf{q} \cdot \mathbf{r}_j(t)}$
and $\mathbf{n}(\mathbf{q};t) = \sum_j \mathbf{n}_j e^{-i \mathbf{q} \cdot \mathbf{r}_j(t)}$.
Next, we follow 
Sec. 8.4 of Hansen and McDonald's monograph \cite{HansenMcDonald} and we 
re-write the first term at the right-hand-side of Eq. (\ref{motion}) as  
\begin{eqnarray}\label{eq:force}
&& i \mathbf{q} \cdot \sum_j \sum_{k \ne j}  \mathbf{r}_{jk} \frac{\mathbf{r}_{jk}}{2 r_{jk}} 
V^{\prime}(r_{jk}) \left[ \frac{e^{i\mathbf{q}\cdot \mathbf{r}_{jk}} - 1}{i \mathbf{q} \cdot \mathbf{r}_{jk}} \right] 
e^{-i \mathbf{q} \cdot \mathbf{r}_j}  \nonumber \\
&=& -i \mathbf{q} \cdot \mathbf{\Pi}_v(\mathbf{q};t), 
\end{eqnarray}
where $\mathbf{r}_{jk} = \mathbf{r}_j - \mathbf{r}_k$ and $\mathbf{\Pi}_v$ has the same form as the virial (interaction) part of 
the pressure tensor. 
We then assume that in the direct space $\mathbf{\Pi}_v(\mathbf{r};t)$ can be expressed in terms of the instantaneous deviation of the 
density from its steady-state average value $\rho$,
\begin{equation}\label{expansion}
\mathbf{\Pi}_v(\mathbf{r};t) \approx \left<\mathbf{\Pi}_v(\mathbf{r};t)\right> 
+ 
\mathbf{I} \left(\partial_\rho P_v\right) \left( \rho(\mathbf{r};t) - \rho \right), 
\end{equation}
where $\mathbf{I}\left(\partial_\rho P_v\right) = \partial \left<\mathbf{\Pi}_v(\mathbf{r};t)\right>/\partial\rho$ and $\mathbf{I}$ 
is the unit tensor.
In the steady state, averages $\left<\mathbf{\Pi}_v(\mathbf{r};t)\right>\equiv P_v$ and $\rho$ are 
translationally invariant. Thus, combining Eqs. (\ref{eq:force}) and 
(\ref{expansion}) we obtain the following approximate expression 
for the first term at the right-hand-side of Eq. (\ref{motion})
\begin{equation}
- i \mathbf{q} \left(\partial_\rho P_v\right)
\sum_j e^{-i \mathbf{q} \cdot \mathbf{r}_j(t)}.
\end{equation}
Next, we take a time derivative and Fourier transform in time and we obtain
\begin{equation}
-\gamma i \omega \mathbf{v}(\mathbf{q};\omega) \approx -\gamma v_o i \omega \mathbf{n}(\mathbf{q};\omega) -
\mathbf{q} \left(\partial_\rho P_v\right) \mathbf{q} \cdot \mathbf{v}(\mathbf{q};\omega).
\end{equation}
To proceed we choose $\mathbf{q} = (q,0)$, which allows us to write
\begin{equation}
v^x(\mathbf{q},\omega) = \frac{i \omega \gamma v_0 n^x(\mathbf{q};\omega)}{\gamma i \omega - q^2 (\partial_\rho P_v)} 
\end{equation}
for the longitudinal correlations. Finally, 
using the same arguments as Henkes \textit{et al.} \cite{Henkes2020}, we obtain the equal-time longitudinal velocity correlations
\begin{equation}\label{corrstheory}
\left< \left|\hat{\mathbf{q}}\cdot\mathbf{v}(\mathbf{q})\right|^2 \right> = \frac{Nv_0^2}{2} \frac{1}{1 + q^2 \tau B_v/(\gamma\rho)}.
\end{equation} 
We identify correlation length $\xi_{\parallel} = \sqrt{\tau B_v/(\gamma\rho)}$ where 
$B_v = \rho \partial_\rho P_v$ is the virial bulk modulus of the active fluid. 

We calculated virial bulk modulus $B_v$ for our active fluid, and found that it depends weakly on $\tau$. $B_v$ increases slightly from
136 for $\tau = 0.14$ to 148 for $\tau = 10.0$ at a fixed $v_0 = 4$ and in our range of persistence times 
it is approximately constant and equal to 145 at fixed $T_a=8$.
The resulting dynamic correlation length
is shown in Figs.~\ref{length} and \ref{lengthT8} as open red symbols. The increase of the correlation 
length $\xi_{\parallel}$ is predominantly due to the increase of the persistence time for both fixed $v_0 = 4$ and fixed $T_a = 8$. 
Our simple theory accurately captures almost all of $\xi_{\parallel}$, but 
it does not predict any transverse velocity correlations. More work is needed to understand these correlations in 
active fluids, as opposed to ordered and amorphous solids discussed in earlier studies.    

\subsection{Properties of our active system}
\begin{figure}
\begin{center}
\includegraphics[width=0.8\columnwidth]{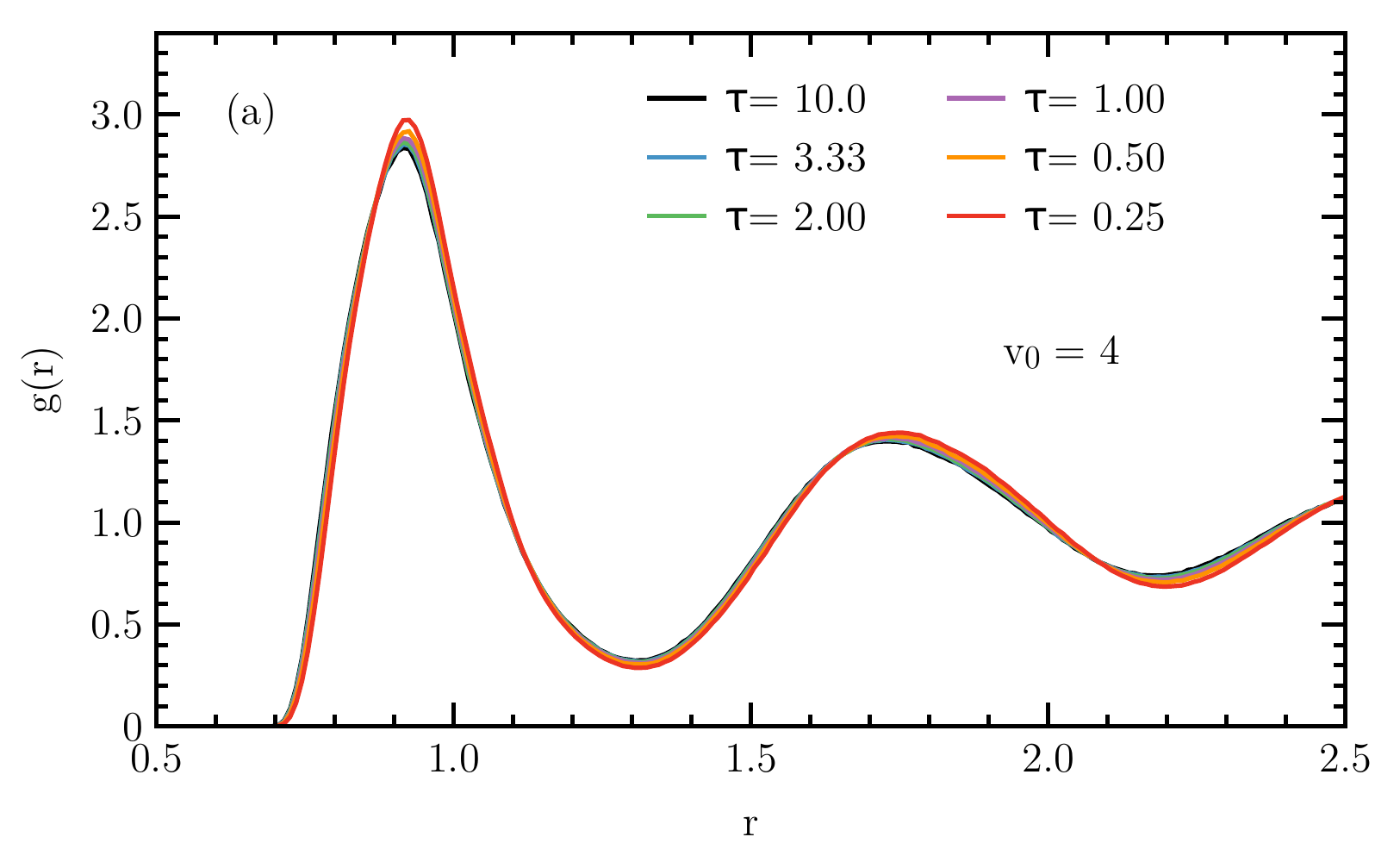}
\includegraphics[width=0.8\columnwidth]{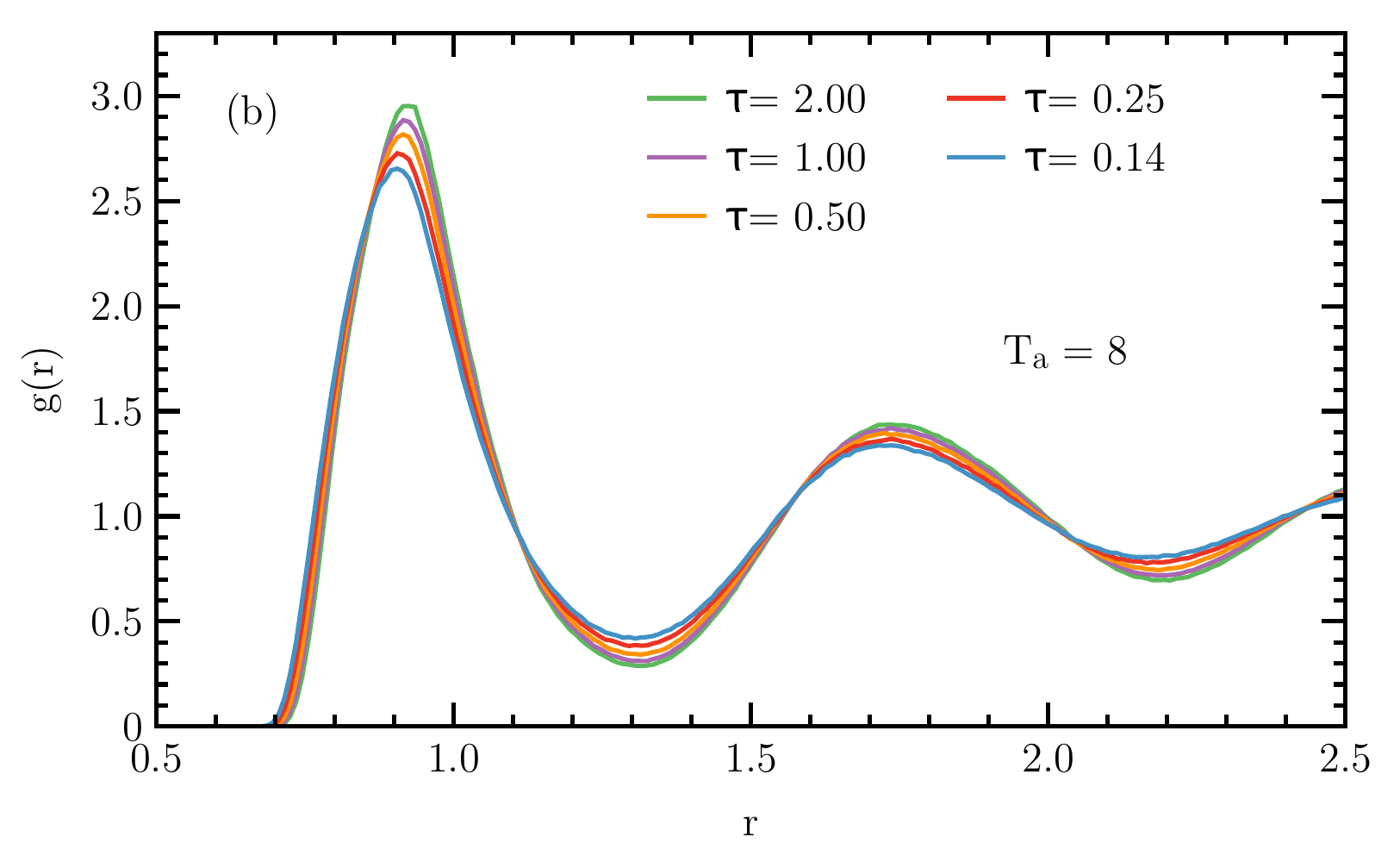}
\end{center}
\caption{\label{gr}Pair correlation functions for (a) fixed $v_0 = 4$ and
(b) fixed $T_a = 8$. There is no significant change in the liquid structure over the full
range of the simulation parameters.}
\end{figure}

In this section we briefly present some of the properties of our system and show that it remains a single-phase fluid in
the range of the parameters that we investigated. First, we evaluated the pair-correlation function 
$g(r) = \frac{1}{\rho N } \left< \sum_{j} \sum_{k \ne j} \delta\left[\mathbf{r} - (\mathbf{r}_j - \mathbf{r}_k) \right]\right>,$
which is sensitive to changes in the local structure and to fractionation that could occur in our polydisperse system. 
Parenthetically, the polydispersity results in significantly broader peaks in $g(r)$ compared to single-component systems.
In Fig.~\ref{gr}(a) we show $g(r)$ for simulations with fixed $v_0 = 4$. We see very little change in the structure 
over the whole range of $\tau$. In general, the height of the first peak decreases slightly with 
increasing $\tau$. In Fig.~\ref{gr}(b) we show $g(r)$ for fixed $T_a = 8$ and, again, we see 
little change from a liquid like structure. In this case, we find that the height of the first peak decreases
with decreasing $\tau$. In both cases we do not see any indication of crystallization or fractionation.

\begin{figure}
\begin{center}
\includegraphics[width=0.8\columnwidth]{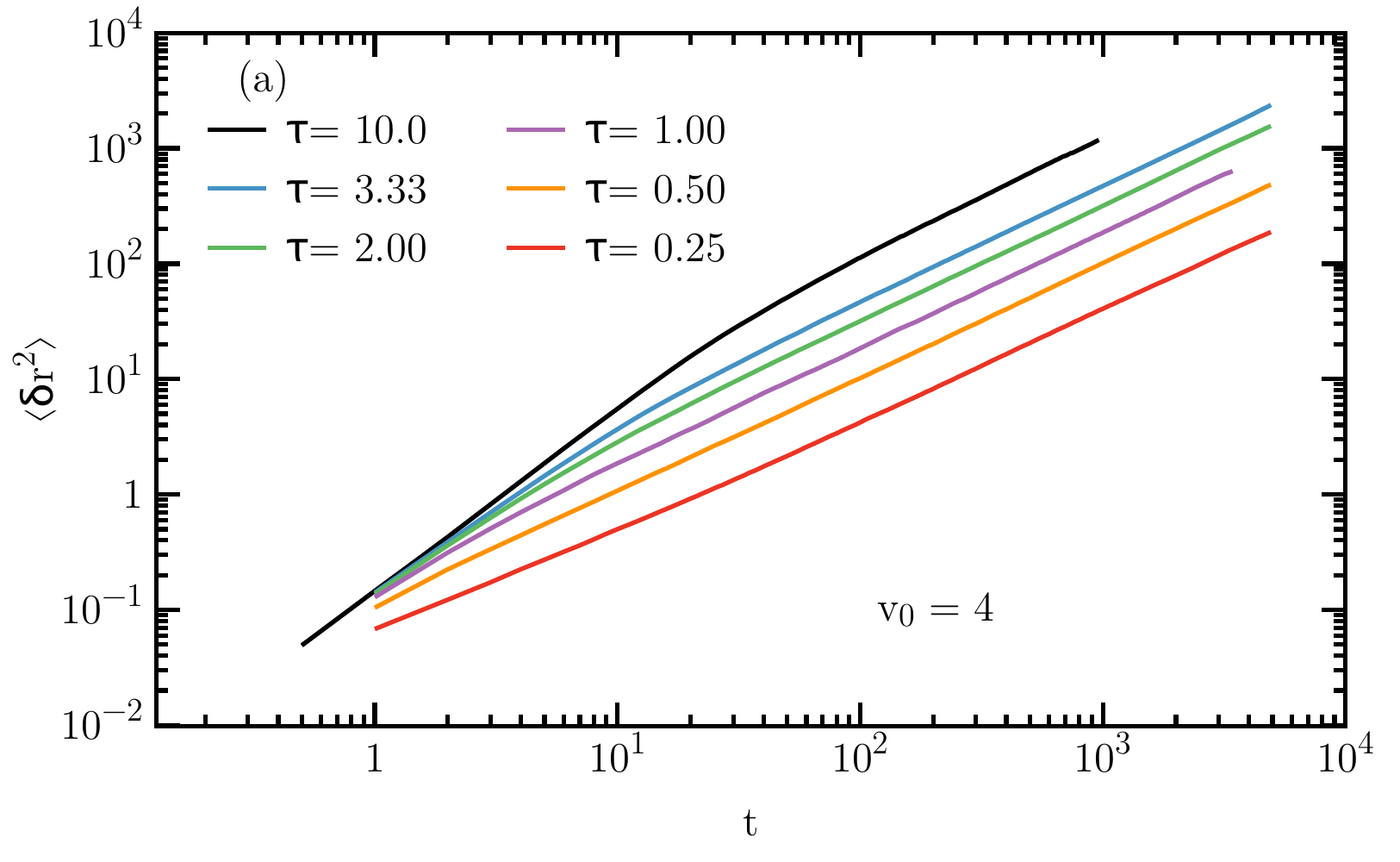}
\includegraphics[width=0.8\columnwidth]{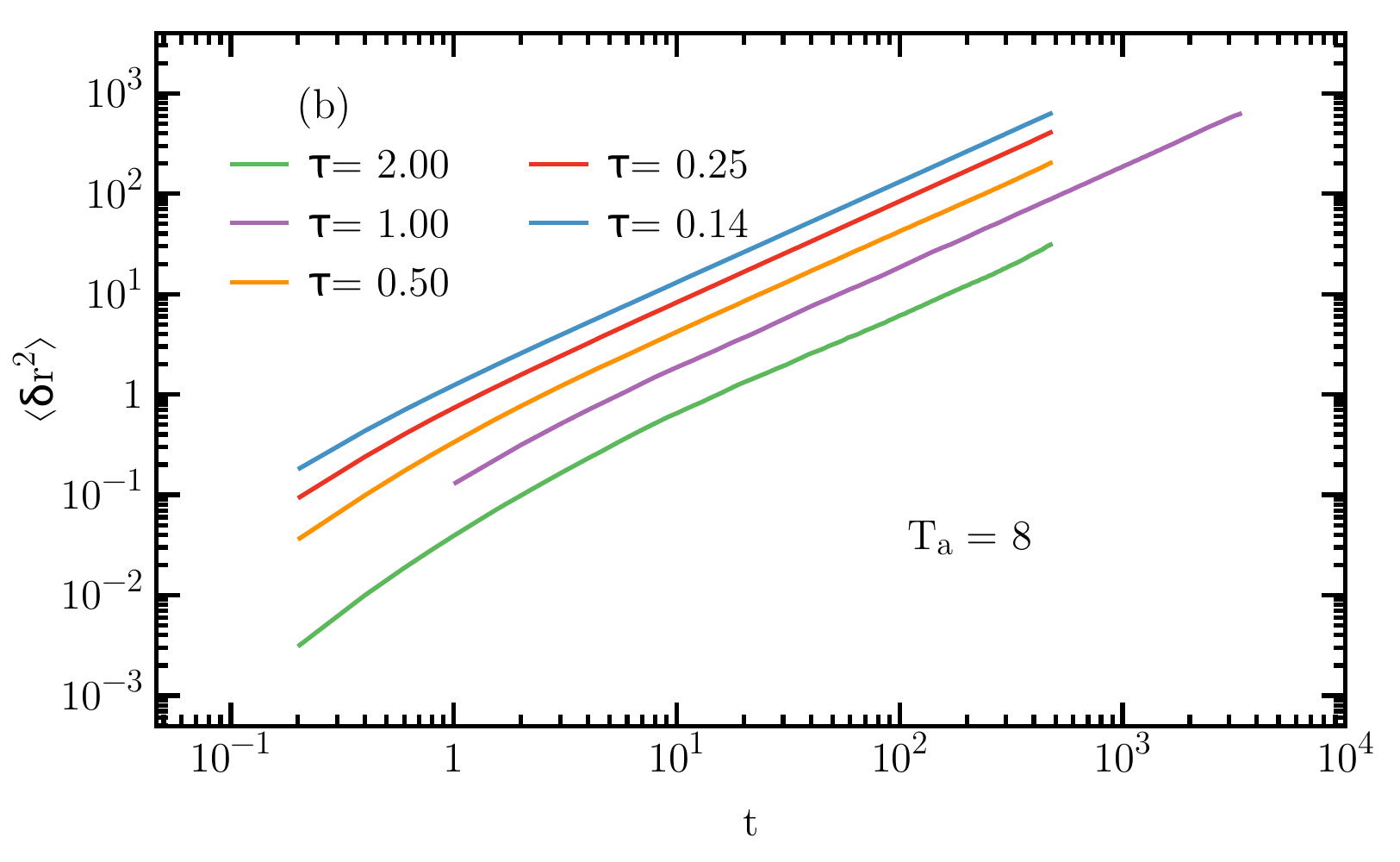}
\end{center}
\caption{\label{msd} The mean-square displacement $\left< \delta r^2(t) \right>$ for 
(a) fixed $v_0 = 4$ and (b) fixed $T_a = 8$ for the range of $\tau$ studied in this work.
All the systems are diffusive at long times and they don't show any signs of arrest or glassy
dynamics. For a fixed $v_0 = 4$ increasing $\tau$ results in faster dynamics, while for 
a fixed $T_a = 8$ increasing $\tau$ results in slower dynamics. Note that for $\tau = 1.0$,
$v_0 = 4$ and $T_a = 8$ correspond to the same state point.}
\end{figure}

To examine the dynamics of our system we evaluated
mean square displacement $\left< \delta r^2(t) \right> = N^{-1} \left< \sum_j [\mathbf{r}_j(t) - \mathbf{r}_j(0)]^2 \right>$.
In Fig.~\ref{msd} we show $\left< \delta r^2(t) \right>$ for the simulations with fixed $v_0=4$ (upper panel) and
for simulations with fixed $T_a = 8$ (lower panel). For both sets of parameters the long time motion is diffusive.
However, the system monotonically speeds up with increasing $\tau$ for fixed $v_0 = 4$, while it monotonically
slows down with increasing $\tau$ for fixed $T_a = 8$. This behavior may be expected for fixed $v_0 = 4$ since the 
diffusion coefficient of an isolated particle for fixed $v_0$ grows as $\tau$. For small enough $\tau$ 
the system may 
approach a structural arrest, which would cause a dramatic slowing down, 
but we do not observe it in these simulations. For fixed $T_a$, on the basis of our earlier work on the dynamics of
systems of self-propelled particles \cite{Berthier2017,Flenner2020}, we would expect a non-monotonic  dependence of the 
long-time diffusion on the persistence time but we did not simulated large enough range of $\tau$ to see it. 

Finally, to check for macroscopic motility-induced phase separation and for the appearance of significant local density fluctuations we 
examined the probability distribution of the local density. To this end we divided the system into squares of length 10
and calculated the probability of the density for these squares. For a system that undergoes motility
induced phase separation one should see this probability bifurcate into high density and low density. The local density
probability distribution that we obtained exhibited a single peak only, indicating single phase, 
and generally changed little in the range of the parameters that we investigated.  

\section{Discussion}
Equal-time velocity correlations are a  ubiquitous feature of active
matter systems, regardless of whether the system is arrested or diffusive and ordered or amorphous. These  
velocity correlations can be very long ranged for large persistence times.
The correlations in the arrested (or almost arrested) systems can be rationalized in terms of the
combined effect of the persistence of the directed motion and the elastic response of these systems. The longitudinal 
correlations in the diffusive systems can be explained in terms of the combined effect of 
the persistence and the fluid's virial bulk modulus, which originates from repulsive interparticle interactions. 
The description of the transverse velocity correlations in
diffusive systems that do not exhibit features of glassy dynamics is an open problem that deserves further study. 

In our opinion, the most interesting open question is the relation of the long-ranged velocity correlations to
the macroscopic properties of active matter. For example, active and passive systems with the 
same local structure, as examined by the pair distribution function or the static structure factor, usually have very different
dynamic properties. Which of the differences can be attributed to the existence of equal-time velocity correlations in
active matter systems? Are these differences sensitive to the range of the velocity correlations? We hope that this
Letter will stimulate further work in this direction.

\acknowledgments
We thank L. Berthier, L. Caprini and A. Shakerpoor for comments on the manuscript. We gratefully acknowledge the support
of NSF Grant No.~CHE 1800282.

\end{document}